\documentclass[twocolumn]{emulateapj}
\usepackage{amssymb}
\begin{document}
\title{FORMATION OF OB ASSOCIATIONS IN GALAXIES}
\author{Frank H. Shu$^1$, Ronald J. Allen$^2$, Susana Lizano$^3$, Daniele Galli$^4$}
\affil{$^1$Department of Physics, University of California, San Diego, CA 92093\\
$^2$Space Telescope Science Institute, Baltimore, MD 21218\\
$^3$CRyA, Universidad Nacional Aut\'onoma de M\'exico, 
Apdo. Postal 72-3, 58089 Morelia, Mexico \\
$^4$INAF-Osservatorio Astrofisico di Arcetri, 
Largo E. Fermi 5, Firenze I-50125, Italy}
\email{fshu@physics.ucsd.edu}
\begin{abstract}
We consider the formation of OB associations from two perspectives: (a) the fractional gas consumption in star formation, $\epsilon$, per dynamical time scale $t_{\rm dyn}$ in a galaxy, and (b) the origin of the so-called Kennicutt-Schmidt law that the rate of star formation per unit area is proportional to a power, $\alpha$, of the surface density in H I and H$_2$ gas when certain thresholds are crossed.  The empirical findings that $\epsilon \approx 10^{-2}$ and $\alpha \approx 1.4$ or $1.5$ have simple explanations if the rate of star formation is magnetically regulated. An empirical test of the ideas resides in an analysis of why giant OB associations are ``strung out like pearls along the arms" of spiral galaxies. 
\end{abstract}

\newcommand{\be}{\begin{equation}}
\newcommand{\ee}{\end{equation}}
\newcommand{\bc}{\begin{center}}
\newcommand{\ec}{\end{center}}

\section{Introduction}

Star formation in the low-redshift universe spans a huge range of linear sizes,
from patterns on tens of kpc in spiral and irregular galaxies, to sub-pc cloud cores collapsing to a small fraction of an AU.  Two empirical rules have brought integration to the subject: (a) the fractional consumption of gas per dynamical (e.g., rotation) time scale $t_{\rm dyn} \sim 10^{8}$ yr is a small number $\epsilon\sim 10^{-2}$, and (b) the rate of star formation per unit area in galaxies is proportional to a nonlinear power of the surface density of cold gas in the galaxy: $\dot \Sigma_* \propto \Sigma_g^{\alpha}$, with $\alpha \approx 1.4$ (Martin \& Kennicutt 2001).

One explanation of these two empirical relations invokes interstellar turbulence and stellar feedback
(Elmegreen \& Scalo 2004, Krumholz \& McKee 2005).  A problem is then the time required to damp turbulent fluctuations, which is $\ell/v$, where $\ell$ is the scale associated with the velocity fluctuation $v$.   This estimate is insensitive to the degree of magnetization of the medium (Stone, Ostriker, \& Gammie 1998; Padoan \& Nordlund 1999).  Prevailing theories postulate the turbulence to be virialized, which implies $\ell/v \sim t_{\rm ff}$, where the free-fall time $t_{\rm ff} \sim t_{\rm dyn} \sim 10^8$ yr for galactic distributions of cold interstellar gas on a scale of a kpc or so.  If turbulence freely decays, the natural star-formation rate on galactic scales would be $\dot \Sigma_* \sim \Sigma_g /t_{\rm dyn}$, $\sim 10^2$ times larger than the observed rate.  Thus, the turbulence scenario faces two difficult obstacles, (a) pumping in energy at a sufficient local rate, usually by artificial means, to offset the natural turbulent dissipation (e.g., MacLow, Klessen, \& Burkert 1998), and (b) explaining an efficiency of star formation $\epsilon\sim 10^{-2}$ on local scales to offset the difficulty if 100\% of the self-gravitating gas were to convert itself into stars on a time scale of $t_{\rm dyn} \sim 10^8$ yr.

\section{The Magnetic Approach}

Fifty years ago Mestel \& Spitzer (1956) noted that star formation would be impossible if the interstellar medium (ISM) were too heavily magnetized.  The natural state of the ISM would then give a rate of star formation which is zero.  The debate between those attracted to magnetism (Mouschovias 1976; Nakano 1979; Shu, Adams, \& Lizano 1987;  McKee 1989) and those favoring turbulence boils down then to whether it is easier to explain $\epsilon = 1$\% starting from 0\% or from 100\%. 

Although the mass-to-flux ratio does not have a well-defined meaning for a galactic disk of gaseous surface density $\Sigma_g$ with an in-plane magnetic field of strength $B$, we introduce the concept of an ISM that is {\it notionally} subcritical in the competition between self-gravitational and magnetic forces:
\be
\lambda_n \equiv {2\pi G^{1/2}\Sigma_g \over B} < 1.
\label{lambda_n}
\ee 
In the solar neighborhood, $\Sigma_g \approx
5 M_\odot$ pc$^{-2}$, $B \approx 5$ $\mu$G, and we easily compute $\lambda_n \approx 0.3$.  Indeed, H I clouds in the Milky Way are all subcritical (Heiles 2004), whereas molecular 
clouds seem marginally critical (Crutcher \& Troland 2006).  

Marginal sub-criticality of the ISM on average is probably no coincidence (Shu et al.~1999).  Heavily supercritical gas clouds have long since collapsed to form stars. To induce the remaining clouds to collapse, any process has to overcome their notional stability. Successful processes only form stars with a low efficiency per unit collapse time scale because magnetic flux on a large scale cannot be destroyed but can only be redistributed.  Thus, gas that becomes supercritical and collapses can only do so by making the remaining gas more subcritical and less prone to collapse.

Under such conditions, the fraction that can collapse to form stars must depend on two time-scales: (a) $t_{\rm dyn}$ of the large-scale dynamical processes that give rise to the proper circumstances for star formation, and (b) the age of the galaxy $t_{\rm gal}$.  Their ratio gives the mean efficiency for star formation: $\epsilon = t_{\rm dyn}/t_{\rm gal} \sim 10^8$ yr/$10^{10}$ yr $\sim$ 1\%.  Galaxies with higher mean efficiencies in the past would have run out of star-forming clouds earlier.  They have become galaxies with lower mean efficiencies today, not classified as spirals, bars, or magellanic-irregulars, but as ellipticals, lenticulars, or anemic spirals.

Unlike the notional $\lambda_n$, the real criterion for gravitational contraction on a large scale to form GMCs involves the surface density $m\bar n(2L)$ if we were to gather gas of mean volumetric number density $\bar n$ and mean molecular mass $m$ along a distance $2L$ in the direction of $\bf B$ (termed the  ``accumulation length'' by Mestel 1985). In the absence of all forces other than self-gravity and electromagnetism, this gathering produces a sheet, with the relevant component of $B_\parallel$ then being the component parallel to the sheet normal (Nakano \& Nakamura 1978, Basu \& Mouschovias 1994, Shu \& Li 1997, Krasnopolsky \& Gammie 2003):
\be
\lambda \equiv {2\pi G^{1/2}m\bar n (2L) \over B_\parallel}.
\label{lambda}
\ee
With orientations as described, $m\bar n = \Sigma_g/2H$, $H$ is the effective half-height
of the gas, $B_\parallel = B$ measured in equation (\ref{lambda_n}),
and $\lambda = \lambda_n L/H$.

This line of thought (e.g., Lizano \& Shu 1989) suggests that there are only two distinct ways by which gas that is notionally subcritical, $\lambda_n < 1$, according to equation (\ref{lambda_n}) can become actually supercritical, $\lambda > 1$, according to equation (\ref{lambda}). Either decrease the denominator of equation (\ref{lambda}) by allowing the magnetic field gas to drift by ambipolar diffusion out of the lightly ionized gas in cold interstellar clouds (e.g., Nakano 1979), or increase the numerator of equation (\ref{lambda})
by gathering the gas along magnetic field lines over a parallel length $2L$ larger than the perpendicular dimension $2H$ to make $\lambda = \lambda_nL/H \ge 1$.

For the birth of massive stars in giant star clusters/associations occurring on dynamical time scales, we focus on the second process, in particular, on  the mechanism of transient gravitational instability in the post-shock regions of the arms of a spiral galaxy (e.g., Shetty \& Ostriker (2006). We wish to extend the analysis to the implications for semi-empirical star-formation laws.

\section{Compression, Contraction, Collapse}

Consider a two-step process beginning with (a) a compression perpendicular to $H$ and $L$ across the field, such as by a magnetized spiral-density-wave shock (Mathewson et al. 1972; Allen, Atherton, \& Tilanus 1986).  The compression (perpendicular to spiral arms) does not change the ratio of $\bar n$ compared to $B_\parallel$ in a thin disk of height $2H$, but it is followed by (b) a periodic contraction along the field (parallel to spiral arms), with expansion in between, which creates the effect of ``feathering" in the gas distribution (see Kim \& Ostriker 2006 and references therein).  Dobbs \& Bonnell (2006) also produce feathering in non-magnetic, non-self-gravitating calculations. The effect is ruined in pure hydrodynamic simulations when one includes the self-gravity of realistic amounts of interstellar gas (P.~R.~Woodward, personal communication in the 1970s), which causes the collapse of entire spiral-arm segments (Chakrabarti, Laughlin, \& Shu 2003).  Sufficient magnetization can prevent this catastrophe (Kim \& Ostriker 2002). 

If the width of perpendicular compression (a) is $W$, the gathering of matter (b) along a length $2L$ in the parallel direction produces a mass $M = 2LW \Sigma_g$. The magnetized flow through the arm makes the gravitational contraction along the length $2L$ a transient instability, a fact we temporarily ignore when we write the equation of motion as
\be
{d^2L\over dt^2} = -{GM \over L^2}.
\label{ODE}
\ee
The solution to equation (\ref{ODE}) gives the time for complete contraction (into a feather) as
\be
t_c = {\pi \over 2\sqrt 2} \left( GM \over L^3\right)^{-1/2} = {\pi\over 4}\left( {G\Sigma_gW\over L^2}\right)^{-1/2},
\ee
with no dependence on $B$ because the contraction is along $\bf B$. Given a post-compression example: $\Sigma_g = 10\; M_\odot$ pc$^{-2}$ (0.5 mag visual extinction for conventional dust-to-gas), $W = 500$ pc for the width of an arm, and $2L = 800$ pc for the half-spacing between giant H II complexes (at the head of feathers) strung along the outer spiral arms of some typical grand-design galaxies (M51, M83, M101), we get $M = 4\times 10^6\, M_\odot$ and $t_c = 7 \times 10^7$ yr, a reasonable mass and age for a GMC complex (cf.~Blitz et al.~2007).  
Contraction of $L$ to  a ``final'' value $< H$ and $W$ produces collapse in 3D. Surviving fluctuations (``turbulence") from the previous spiral-arm crossing introduces clumpiness that prevents actual complete collapse.

If $H = 100$ pc, and $\lambda_n = 0.3$, as in the solar neighborhood before and after perpendicular compression, then $\lambda = \lambda_n L/H = 1.2$ after parallel contraction. Presumably, $L$ is set by the time available locally to make $\lambda > 1$ marginally. Similarly, $t_c$ computed from observational data is comparable to the rotation time scale $t_{\rm dyn}$ because, apart from magnetism, the feathering instability has to overcome the tidal forces of the galactic gravitational field, the shear and rotation across spiral arms, and any turbulent velocities.  This threshold is governed by some generalization of Toomre's (1964) $Q$ parameter.  

With a resultant $\lambda = \lambda_n L/H$ greater than unity, the rate of GMC formation per unit area, $\dot \Sigma_{\rm GMC}$, is given by $\Sigma_g/t_c$ times the mass fraction $\epsilon_{\rm GMC}$ of total gas converted into GMCs:
\be
\dot \Sigma_{\rm GMC} = \epsilon_{\rm GMC} {\Sigma_g \over t_c} = {4 \over \pi}\epsilon_{\rm GMC}  \left({GW\over L^2}\right)^{1/2}\Sigma_g^{3/2}.
\label{GMCrate}
\ee
Because half the gas is between feathers, we estimate $\epsilon_{\rm GMC} \sim 1/2$.

Being a mixture of supercritical cores and subcritical envelopes initiated by the turbulence of the recurring feathering instability, GMCs do not form stars with unit efficiency. The diverging mean flow in the interarm region causes attached magnetic fields to tear the envelopes apart in the $W$ direction faster than they can continue to accumulate mass in the $L$ direction.  To obtain the rate of star formation per unit area $\dot \Sigma_*$ associated with GMC formation, we must multiply equation (\ref{GMCrate}) by the mass fraction of gravitationally bound, supercritical cores that goes into continued gravitational collapse, $\epsilon_{\rm cores}$, times the mass fraction of collapsing cores that results in the formation of stars, $\epsilon_{\rm SF} \sim 1/3 $ (Alves, Lombardi, \& Lada 2007):  
\be
\dot \Sigma_* = \epsilon_{\rm cores}\epsilon_{\rm SF}\dot \Sigma_{\rm GMC},
\label{sfrate}
\ee
Empirically, $\epsilon_{\rm cores}$ ranges from 1\% (Alves et al.~2007) to 10\% or more (Alves, Lada, \& Lada 1999); we adopt $\epsilon_{\rm cores} \sim 6\%$ as a representative value. Combining equations (\ref{GMCrate}) and (\ref{sfrate} gives
a Kennicutt-Schmidt law, $\dot \Sigma_* \propto \Sigma_g^{\alpha}$ with $\alpha = 3/2$ and a coefficient $\propto$ total efficiency $\epsilon = \epsilon_{\rm GMC}\epsilon_{\rm cores}\epsilon_{\rm SF}$ estimated as $(1/2)(6\%)(1/3) = 1\%$. This estimate has large local scatter, but the argument of ``natural selection'' of \S 2 -- that star formation on a galactic scale converts, in the mean, a significant fraction of the available gas into stars only on a time scale equal to the age of the system   -- ensures that $\epsilon \sim 1$\% as a global average.

For our conclusion on $\alpha$ to hold, $W^{1/2}/L$ must have little dependence on $\Sigma_g$.  The width $W$ of gaseous spiral arms is set primarily by the properties of the background, {\it stellar}, spiral density-wave and not by $\Sigma_g$ (Roberts \& Yuan 1970).  The accumulation length $2L$ is set, we have already claimed, by the need to make the post-shock $\lambda$ marginally greater than 1. Observationally, $L$ has much less variation from region to region than $\Sigma_g$. And the large-scale turbulence at $\sim 8$ km s$^{-1}$, renewed at each spiral-arm crossing, can have a decay time $\sim 800 \; {\rm pc /8\; \rm km \; s}^{-1} = 10^8\, {\rm yr}$.

The results (\ref{GMCrate}) and (\ref{sfrate}) are independent of the scenario that motivated their derivation. Similar arguments should apply to the weak B-star formation that occurs in GALEX images of outer disks (Boissier et al.~2007) or to the``starbursts'' in interacting galaxies like the Antennae system, NGC 4038/9 (Zhang, Fall, \& Whitmore 2001). Nevertheless, the development of a rigorous theory of the process and observational tests of its predictions are most decisively carried out by a resurrection of the classical problem of why giant H II regions in a spiral or barred galaxy are ``strung out like pearls along the arms" (Baade 1963).

\bigskip\noindent
We are grateful to Art Wolfe for discussions and to Eve Ostriker
for her illuminating papers with Kim and Shetty, which inspired this work. The questions raised by an expert referee helped greatly to sharpen the final arguments.

{}
\end{document}